\documentclass[twoside,twocolumn,9pt]{article}
\usepackage{extsizes}
\usepackage[super,sort&compress,comma]{natbib} 
\usepackage[version=3]{mhchem}
\usepackage[left=1.5cm, right=1.5cm, top=1.785cm, bottom=2.0cm]{geometry}
\usepackage{balance}
\usepackage{mathptmx}
\usepackage{sectsty}
\usepackage{graphicx} 
\usepackage{lastpage}
\usepackage[format=plain,justification=justified,singlelinecheck=false,font={stretch=1.125,small,sf},labelfont=bf,labelsep=space]{caption}
\usepackage{float}
\usepackage{fancyhdr}
\usepackage{fnpos}
\usepackage[english]{babel}
\addto{\captionsenglish}{%
  
}
\usepackage{array}
\usepackage{droidsans}
\usepackage{charter}
\usepackage[T1]{fontenc}
\usepackage[usenames,dvipsnames]{xcolor}
\usepackage{setspace}
\usepackage[compact]{titlesec}
\usepackage{hyperref}

\usepackage{epstopdf}

\definecolor{cream}{RGB}{222,217,201}

\begin{document}

\pagestyle{fancy}
\thispagestyle{plain}
\fancypagestyle{plain}{
\renewcommand{\headrulewidth}{0pt}
}

\makeFNbottom
\makeatletter
\renewcommand\LARGE{\@setfontsize\LARGE{15pt}{17}}
\renewcommand\Large{\@setfontsize\Large{12pt}{14}}
\renewcommand\large{\@setfontsize\large{10pt}{12}}
\renewcommand\footnotesize{\@setfontsize\footnotesize{7pt}{10}}
\makeatother

\renewcommand{\thefootnote}{\fnsymbol{footnote}}
\renewcommand\footnoterule{\vspace*{1pt}%
\color{cream}\hrule width 3.5in height 0.4pt \color{black}\vspace*{5pt}} 
\setcounter{secnumdepth}{5}

\makeatletter 
\renewcommand\@biblabel[1]{#1}            
\renewcommand\@makefntext[1]%
{\noindent\makebox[0pt][r]{\@thefnmark\,}#1}
\makeatother 
\renewcommand{\figurename}{\small{Fig.}~}
\sectionfont{\sffamily\Large}
\subsectionfont{\normalsize}
\subsubsectionfont{\bf}
\setstretch{1.125} 
\setlength{\skip\footins}{0.8cm}
\setlength{\footnotesep}{0.25cm}
\setlength{\jot}{10pt}
\titlespacing*{\section}{0pt}{4pt}{4pt}
\titlespacing*{\subsection}{0pt}{15pt}{1pt}

\fancyfoot{}
\fancyfoot[LO,RE]{\vspace{-7.1pt}\includegraphics[height=9pt]{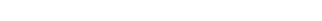}}
\fancyfoot[CO]{\vspace{-7.1pt}\hspace{13.2cm}\includegraphics{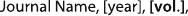}}
\fancyfoot[CE]{\vspace{-7.2pt}\hspace{-14.2cm}\includegraphics{head_foot/RF}}
\fancyfoot[RO]{\footnotesize{\sffamily{1--\pageref{LastPage} ~\textbar  \hspace{2pt}\thepage}}}
\fancyfoot[LE]{\footnotesize{\sffamily{\thepage~\textbar\hspace{3.45cm} 1--\pageref{LastPage}}}}
\fancyhead{}
\renewcommand{\headrulewidth}{0pt} 
\renewcommand{\footrulewidth}{0pt}
\setlength{\arrayrulewidth}{1pt}
\setlength{\columnsep}{6.5mm}
\setlength\bibsep{1pt}

\makeatletter 
\newlength{\figrulesep} 
\setlength{\figrulesep}{0.5\textfloatsep} 

\newcommand{\topfigrule}{\vspace*{-1pt}%
\noindent{\color{cream}\rule[-\figrulesep]{\columnwidth}{1.5pt}} }

\newcommand{\botfigrule}{\vspace*{-2pt}%
\noindent{\color{cream}\rule[\figrulesep]{\columnwidth}{1.5pt}} }

\newcommand{\dblfigrule}{\vspace*{-1pt}%
\noindent{\color{cream}\rule[-\figrulesep]{\textwidth}{1.5pt}} }

\makeatother

\twocolumn[
  \begin{@twocolumnfalse}
{\includegraphics[height=30pt]{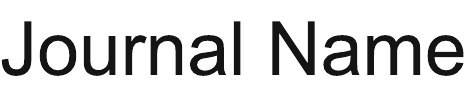}\hfill\raisebox{0pt}[0pt][0pt]{\includegraphics[height=55pt]{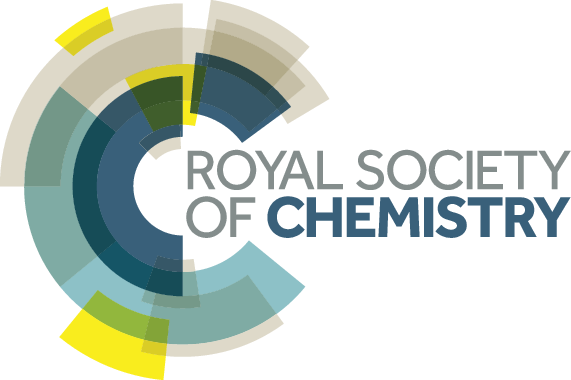}}\\[1ex]
\includegraphics[width=18.5cm]{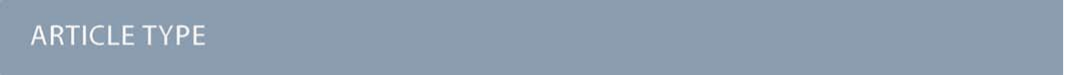}}\par
\vspace{1em}
\sffamily
\begin{tabular}{m{4.5cm} p{13.5cm} }

\includegraphics{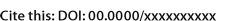} & \noindent\LARGE{\textbf{Indirect-to-direct bandgap transition in few-layer $\beta$-InSe as probed by photoluminescence spectroscopy}} \\
\vspace{0.3cm} & \vspace{0.3cm} \\

 & \noindent\large{Bogdan R. Borodin, Ilya A. Eliseyev, Aidar I. Galimov, Lyubov V. Kotova, Mikhail V. Durnev, Tatiana V. Shubina, and Maxim V. Rakhlin$^{\ast}$} \\

\includegraphics{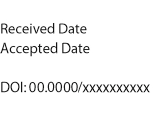} & \noindent\normalsize{InSe is a promising material for a next-generation of two-dimensional electronic and optical devices, characteristics of which are largely determined by the type of band structure, direct or indirect. In general, different methods can be sensitive to different peculiarities of the electronic structure leading to different results. In this work, we will focus on the luminescent properties of few-layer $\beta$-InSe with a thickness of 6 to 75 monolayers (ML). Low-temperature micro-photoluminescence ($mu$-PL) studies show a sharp increase in PL intensity in the range of thicknesses from 16 to 20 monolayers, where, in addition, there is a singularity in the dependence of the work function on the thickness. Time-resolved photoluminescence spectroscopy (TRPL) reveals three characteristic PL decay times that differ from each other by about an order of magnitude. We associate the processes underlying the two faster decays with the recombination of electrons and holes between the band extrema, either directly or through the interband relaxation of holes. Their contributions to the total PL intensity increase significantly in the same thickness range, 16-20 MLs. On the contrary, the slowest contribution, which we attribute mainly to the defect-assisted recombination,
prevails at a smaller number of monolayers and then noticeably decreases. 
These results indicate the indirect-to-direct bandgap transition near 16-20 MLs, which determines the range of applicability of a few-layer $\beta$-InSe for efficient light emitters.} 

\end{tabular}

 \end{@twocolumnfalse} \vspace{0.6cm}

  ]

\renewcommand*\rmdefault{bch}\normalfont\upshape
\rmfamily
\section*{}
\vspace{-1cm}


\footnotetext{\textit{Ioffe Institute, St. Petersburg, 194021, Russia; E-mail: maximrakhlin@mail.ru}}





\section{Introduction}

The booming development of the field of two-dimensional (2D) materials has led to the emergence of several classes of layered materials designed for various applications.~\cite{mas20112d,Geim,Chaves:2020,Liu:2021,Montblanch:2023}. Among them, the  post-transition metal chalcogenides (PTMCs), such as InSe and GaSe, are very promising from both fundamental and applied perspective. ~\cite{ho2017high,late2012gas,lu2020multilayer,chen2020high}
In contrast to transition metal dichalcogenides\cite{mak2010atomically,Wang:2018}, PTMCs monolayers have an indirect band gap that transforms to direct with increasing thickness. This transition is related to the change of the valence band dispersion: Parabolic dispersion specific to bulk PTMCs turns into “Mexican hat” dispersion in monolayers~\cite{zolyomi2013band,zolyomi2014electrons,rybkovskiy2014transition,guo2017band,sun2018inse}.

The “Mexican hat” dispersion of holes with an extremum loop results in many unique electronic properties and physical phenomena in PTMCs.~\cite{cao2015tunable, lugovskoi2019strong, liu2022pressure,chen2019phonon,chen2022emergence, iordanidou2018hole}. Among PTMC materials, InSe is the most actively studied as it has many possible applications in high-mobility electronic devices\cite{arora2021recent,ho2017high,tsai2019high,late2012gas}, spintronics~\cite{iordanidou2018hole,premasiri2018tuning,liu2018graphene}, optoelectronics~\cite{wang2020reduction,lu2020multilayer,chen2018ultrafast}, flexible devices~\cite{chen2020high,zhang2022fully,wu2020inse}, and straintronics~\cite{ibarra2022modification,maeso2019strong,li2018ultrasensitive,wang2020scaling}.
However, despite significant progress in the fundamental and application-related studies of InSe, the optical properties of a few-layer InSe are still poorly understood. Of particular interest is the transition between indirect and direct band gaps and its influence on the photoluminescence (PL) of InSe layers.

Theoretical works predict a smooth transformation of the band gap from indirect to direct as the number of layers increases from 1 to about 20. \cite{sang2019two,sun2018inse,rybkovskiy2014transition,iordanidou2018hole}. Meanwhile, experimental data are controversial. Studies of the optical properties show different PL dynamics with increasing thickness and the presence of PL starting from already 2 monolayers (ML)~\cite{song2020optical,Mudd:2016,Bandurin:2017,Mudd:2013,Zheng:2017,venanzi2020}. The study of $\gamma$-InSe band structure evolution using angle-resolved photoemission spectroscopy by Hamer et al. indicates that, after the thickness of 3 ML, the band inversion is no longer measurable as it is less than $kT$ at room temperature\cite{hamer2019indirect}. There are several reasons for these discrepancies. The first reason is the presence of four different InSe polytypes ($\gamma$-, $\epsilon$-, $\beta$- and $\delta$-phases), which have different crystallographic structures and, hence, slightly different electronic and optical properties.
The vast majority of experimental works are dedicated to $\gamma$-InSe \cite{Mudd:2013, Mudd:2016, Bandurin:2017}, while a part of works does not specify the polytype \cite{Zheng:2017,venanzi2020}. On the other hand, few-layer $\beta$-InSe is much less studied despite the unique excitonic properties of the bulk material~\cite{Shubina2019}.
 The second reason is the non-cryogenic temperatures of most experiments. According to theoretical calculations\cite{sang2019two,sun2018inse,rybkovskiy2014transition,iordanidou2018hole, zolyomi2014electrons, Magorrian2016}, the valence band changes smoothly in the transition region, and hence,
the high $kT$ energy can blur the moment of transition. The third reason is the natural oxidation of InSe in ambient conditions which creates different types of defects involved in optical and electronic processes\cite{xiao2017defects,yang2020oxidation,zhang2021enhancement,tsai2019high,Shubina2019,venanzi2020}.

In general, different methods can be sensitive to different aspects of the electronic structure and therefore can detect the transition from indirect to direct bandgap at different numbers of monolayers for a given polytype, which is an important factor for designing optoelectronic devices. In this work, we use PL spectroscopy at cryogenic temperature to detect the transition from indirect to direct bandgap in a few-layer $\beta$-InSe, which was first studied by a Raman spectroscopy, atomic force microscopy (AFM), and Kelvin probe force microscopy (KPFM) that showed a singularity in the dependence of work function at a flake thickness of 16–20 ML. Micro-photoluminescence ($\mu$-PL) demonstrated a sharp increase in the PL intensity in flakes of the same thickness range, which we explained by the dominance of direct optical transitions. Additional information was obtained using time-resolved photoluminescence (TRPL), which shows three characteristic PL decay times that differ from each other by about an order of magnitude. We attribute the processes underlying the two faster PL decays to direct-band recombination of electrons and holes (either free or bound in exciton), which occurs either immediately between band extrema or with the participation of interband hole relaxation. The slowest decay is due to a combination of defect-assisted and possibly indirect optical transitions; the latter component disappears when the direct type of the band structure is finally established. The simultaneous increase in the contributions of two faster direct transitions emphasizes the transformation of the valence band from the "Mexican hat" to the parabolic one, probed by PL.


\section{Samples}
The structures under study were fabricated from a commercially available high quality InSe bulk crystal (HQ-graphene production). For sample preparation, the so-called Scotch tape method was used. 

\begin{figure}[h]

 \includegraphics[width=0.45\textwidth]{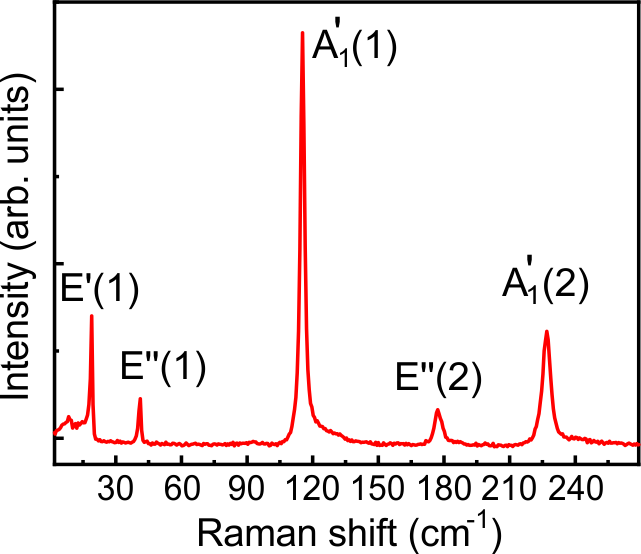}

   \caption{Raman spectrum of the bulk InSe crystal used for exfoliation}.
   \label{Raman}
\end{figure}

The structural quality of the samples was confirmed by micro-Raman studies. Figure \ref{Raman} presents the Raman spectra of bulk InSe crystal used for the preparation of few-layer flakes. The spectrum represents lines corresponding to phonons in InSe: $E^{'}(1)$ (19 cm$^{-1}$), $E^{''}(1)$ (41 cm$^{-1}$), $A^{'}(1)$ (115 cm$^{-1}$), $E^{''}(2)$ (177 cm$^{-1}$), and $A^{'}(2)$ (227 cm$^{-1}$). We did not observe the Raman line at 199 cm$^{-1}$ corresponding to the $E^{'}(TO)$ phonon, which is known to become Raman-active only in non-centrosymmetric polytypes, such as $\epsilon$ or $\gamma$ \cite{jandl1978}. Thus, we may attribute the bulk material and hence the exfoliated flakes to the $\beta$ polytype. 

\begin{figure*}
  \includegraphics[width=1\textwidth]{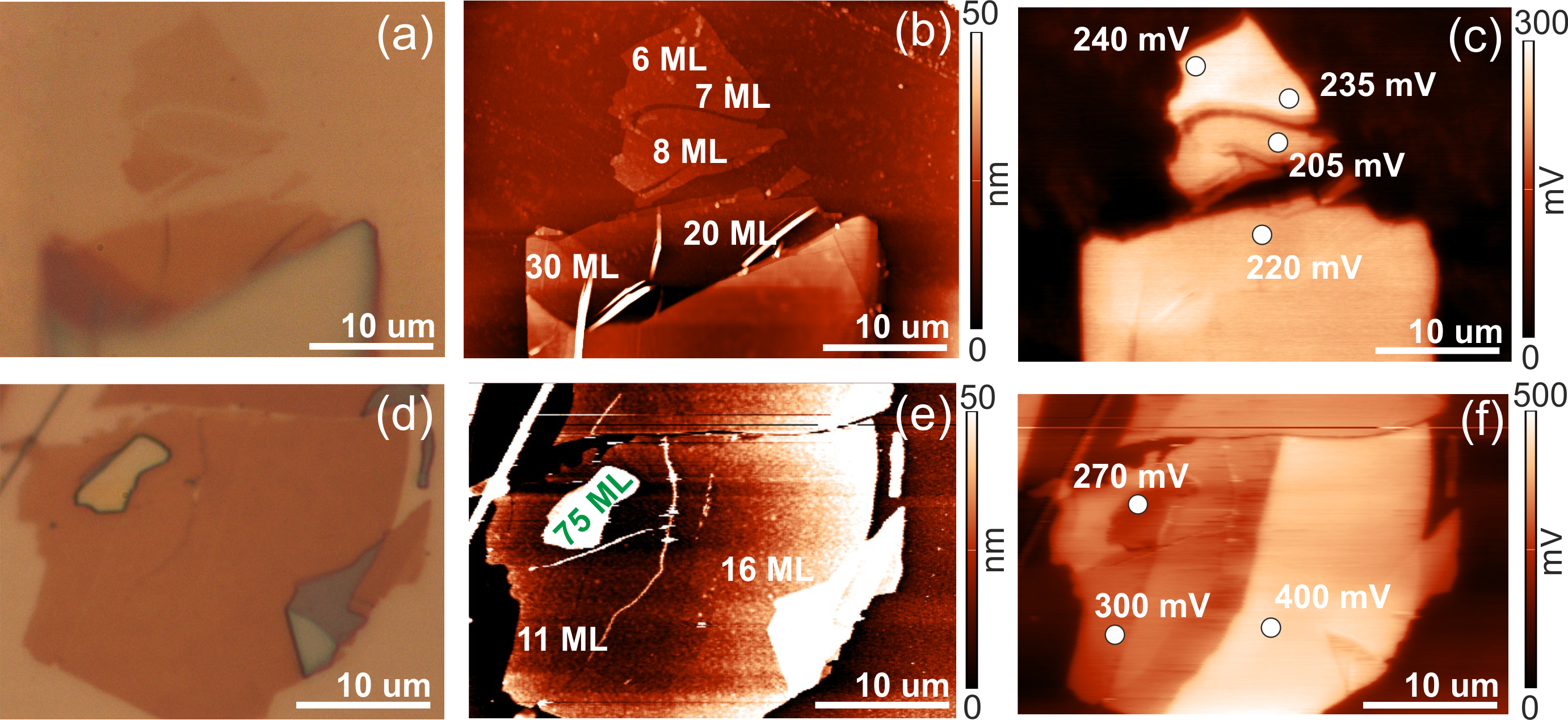}
  \caption{Optical images (a, d), topography (b, e) and surface potential distribution (c, f) of the selected flakes.}
   \label{Topo}
\end{figure*}

After fabrication, the topography of the samples was    studied using atomic force microscopy. Several flakes were selected based on their thickness. 
We picked flakes with thicknesses ranging from 6 to 75 to investigate the evolution of electronic and optical properties of a few-layer $\beta$-InSe. The optical images of the flakes are shown in Figs.~\ref{Topo}(a, d), while Figs.~\ref{Topo}(b, e) demonstrate the topography of the selected flakes with the indicated number of layers as it was measured by SEM.
As seen from Fig.~\ref{Topo}, the obtained flakes are sufficiently large, which makes it possible to obtain reliable data on  PL  and surface potential. Figures~\ref{Topo}(c, f) show the surface potential distribution maps. 
We can see that the surface potential depends non-monotonously on the thickness and even experiences leaps between pairs of flakes of a certain thickness, while random jumps of 5 mV and 30 mV between, respectively, flakes 6 ML-7 ML and 7 ML-8 ML may reflect the influence of the surface or substrate, a jump of 180 mV between flakes 16 ML and 20 ML indicates a cause possibly related to the transformation of band structure. We believe that if such a rapid change in the surface potential indicates this transformation, then it should also manifest itself in the optical properties. To confirm this assumption, we studied the flakes PL, since both its intensity and characteristic decay times should depend on the type of band gap, direct or indirect.

\section{Emission properties}

\subsection{Micro-photoluminescence}

\begin{figure} [h]
\centering
  \includegraphics[width=0.48\textwidth]{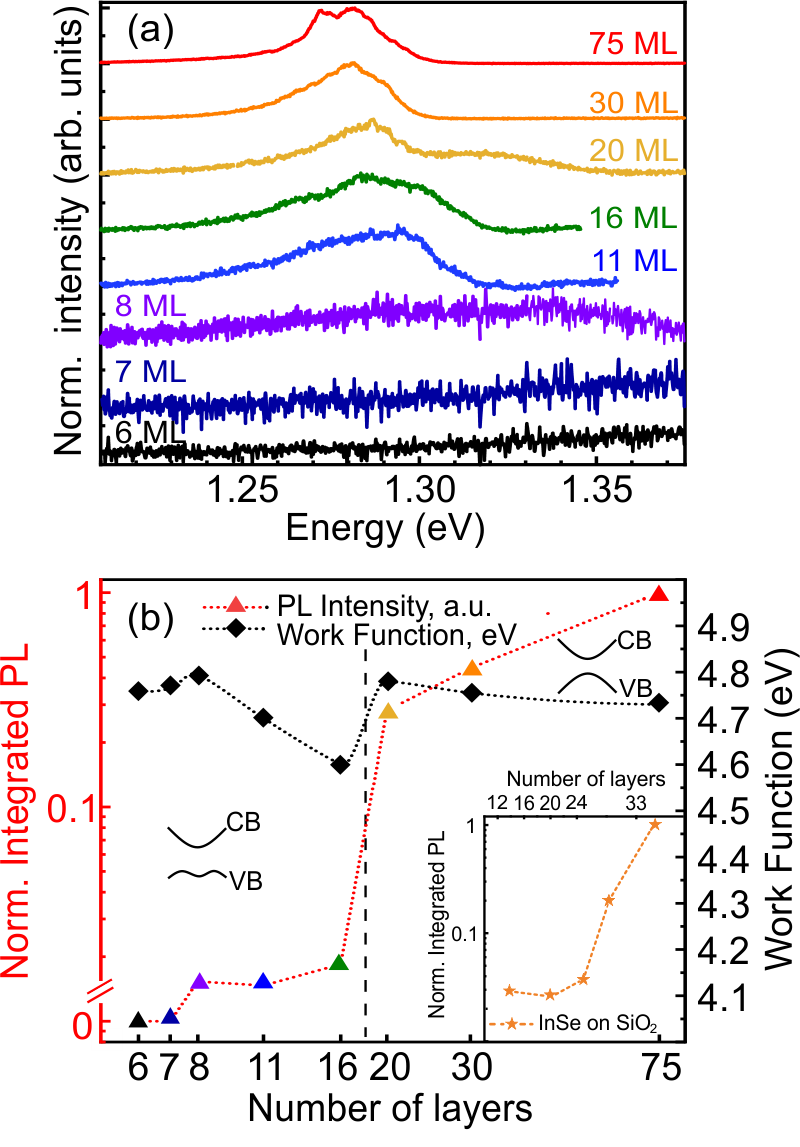}
  \caption{(a) PL spectra of the flakes with different thickness at 4 K. Each spectrum is normalized to its maximum and shifted vertically for clarity. (b) Integrated PL intensity normalized to its value at 75 ML and the work function in flakes with different thickness. The black dashed line shows an approximate area of indirect-to-direct bandgap transition. The insets show normalized integrated PL intensity of InSe on SiO$_2$.}
   \label{PL}
\end{figure}

Figure~\ref{PL} (a) demonstrates the PL spectra of the flakes obtained at 4 K.
One can see that both the spectral behaviour and the PL intensity (characterized by a noise track) strongly depend on the InSe thickness. The integrated PL intensity and the work function vs. the number of layers are summarized in Fig.~\ref{PL} (b). 6- and 7-ML flakes do not show any detectable PL signal. PL appears in a 8 ML flake and barely increases up to 16-ML thickness. In a 20 ML flake, PL intensity rapidly increases and then linearly grows from 20 to 75 MLs. We also observed a similar behavior of the PL intensity on InSe flakes on SiO$_2$ (see inset in  Fig.~\ref{PL} (b)). We believe that the absence of PL for thin flakes ($<$ 8 ML) is associated with the dominant role of surface defects, which are common for InSe exposed to ambient conditions\cite{hopkinson2019formation}. These defects provide a non-radiative recombination channel. Also, the influence of the metal when using Au/Si substrates cannot be ruled out. With the thickness increase ($\geq$ 8 ML), the role of volume increases, and weak PL appears.
As described above, the work function behaves non-monotonously with the number of layers and varies between 4.6 to 4.8 eV. Its sharp decrease occurs at about 20 mV, and then an increase follows with a decrease in the number of layers, probably due to enhanced size quantization \cite{sang2019two}. The singuliarity at 16-20 ML agrees well in terms of the layer number with a sharp increase in the PL intensity.

Superlinear growth of the PL intensity with increasing thickness has been previously observed in InSe flakes of different polytypes~\cite{Mudd:2013,Mudd:2016,Bandurin:2017,Zheng:2017,venanzi2020} and has been attributed to the transition between indirect and direct band gap in InSe. Density functional theory (DFT) calculations predict the valence-band dispersion in InSe to change from the Mexican-hat-shaped (or ring-shaped) to the parabolic-shaped with increasing number of layers~\cite{rybkovskiy2014transition,zolyomi2014electrons,Magorrian2016}. For the “Mexican hat” valence-band configuration expected for InSe, for layer thicknesses from one to several MLs, the band gap is indirect. Based on the analysis of the steady-state and time-resolved (discussed below) PL, we conclude that the flakes with $N \leq 16$ are indirect-band. Hence, we attribute the rapid increase of PL intensity in the 16-20 ML interval, see Fig.~\ref{PL} (b), to the transformation of the valence band and indirect-to-direct band gap transition. The obtained results are consistent with the theoretical calculations obtained by Rybkovskiy et al. \cite{rybkovskiy2014transition} The smooth PL growth in the 8-16 and 20-75 ML intervals is associated with an increase of the emitting volume. Note that the work function leap between 16 and 20 ML correlates with the rapid increase of PL in the same flakes.

\subsection{Time-resolved photoluminescence}

The broad low-temperature PL spectra shown in Fig.~\ref{PL} (a) indicate a possible contribution of the defect-assisted radiative recombination, which has been mentioned in previous optical studies of a few-layer and bulk InSe~\cite{Mudd:2013,Shubina2019,venanzi2020}. To gain a further insight in the nature of PL, we use the TRPL spectroscopy. As far as we know, only Venanzi et al. have experimentally determined the decay times in InSe\cite{venanzi2020}, but the measurement range was insufficient to accurately determine the long components of the decay curves, so we used a laser with 4 MHz freaquency to refine their data. Figure~\ref{TRPL} shows the results of TRPL measurements in four flakes of 11, 16, 20 and 75 MLs. For all the studied thicknesses, the best fit for the decay curve was achieved using 3 exponential components, which were considered as corresponding to the PL itself, convoluted with the instrumental function (for procedure description, see \cite{eliseyev2021mos2}). The function used for the fit has the form
\begin{equation}
    f(t) = i_1 \mathrm{e}^{-t/\tau_1} + i_2 \mathrm{e}^{-t/\tau_2} + i_3 \mathrm{e}^{-t/\tau_3}\:,
\end{equation}
where $\tau_j$ are the characteristic of decay times, and $i_j$ are the corresponding PL intensities, $j = 1,2,3$. The total (time-integrated) intensity of the $j$-th component is given by $I_j = i_j \tau_j$. Figures~\ref{TRPL} (b) and (c) show the relative intensities of each component given by $I_j/I$, where $I = I_1 + I_2 + I_3$, and the decay times $\tau_j$ as functions of the number of layers.
 
\subsection{Optical transitions}

\begin{figure*}
\centering
   \includegraphics[width=1\textwidth]{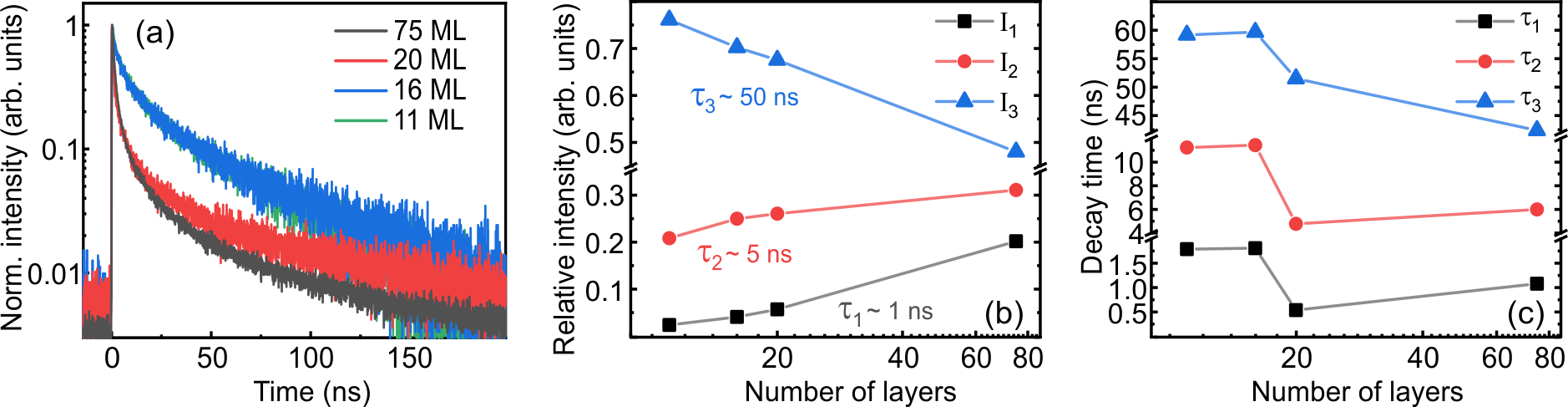}
    \caption{(a) Normalized PL decay curves measured from four InSe flakes with different thicknesses (11, 16, 20 and 75 ML). 
    (b) Relative intensities of the “fast” ($I_1/I$), “medium” ($I_2/I$) and “slow” ($I_3/I$) components of PL as a function of number of layers. Corresponding decay times are marked on the graph. (c) Decay times $\tau_1$, $\tau_2$ and $\tau_3$ as a function of the number of layers.}
    \label{TRPL}
\end{figure*}

The first component of the PL decay, referred to as “fast”, has a characteristic decay time $\tau_1 \sim 0.5-1$~ns, the second, “medium”, component decays at the time scale of $\tau_2 \sim 4-11$~ns, and the third, “slow”, component decays in $\tau_3 \sim 40-60$~ns, see Fig.~\ref{TRPL} (c).
We attribute part of the “slow” component to the defect-related  recombination, as mentioned in previous studies of thin-layer InSe ~\cite{venanzi2020}. In this case, an electron or/and a hole are localized at defects, which results in a small overlap of their wave functions in the coordinate or momentum space. This, in turn, leads to a small radiative recombination rate, and consequently, large decay time $\tau_3$. The line associated with the defects is located somewhat below the emission of the free exciton, and during measuring the TRPL, we must inevitably register a part of that. The other part of the “slow” component might be ascribed to the indirect exciton transition situated a bit above. As any indirect transition it is slow especially at low temperature when the phonon population is poor. While the defect-related contribution is weakly dependent on the flake thickness, the fundamental indirect part should strongly follow the indirect-to-direct transition. Our measuring the few-layer InSe is in line with that.
As shown in Fig.~\ref{TRPL} (b), the relative contribution of the defect-assisted processes to the total PL ($I_3/I$) is large, and, in fact, dominates the PL at $N \leq 20$. However, as the critical thickness of 16–20 ML is approached, the contribution of the “slow” component decreases. The remainder is due to defect-related radiation captured due to our measurement conditions. At the same time, a gradual increase in two “faster” components is observed, the presence of which, as will be described below, is due to the specific band structure of InSe.

We assume that the “faster” component of the PL decay ($I_1$ and $\tau_1$) is due to the direct band-to-band recombination of electrons and holes (either free or bound in exciton). The decay rate $\tau_1^{-1}$ includes both radiative and non-radiative processes, the latter being, e.g., the capture of a free electron/hole to a defect. The radiative lifetime $\tau_r$ can be estimated from $\tau_r^{-1} = 4 \omega^3 d_z^2/(3\hbar c^3)$, where $\omega$ is the transition frequency, $d_z$ is the matrix element of the electric dipole moment between the conduction and valence band states at $\bf{k}$ = 0, $\hbar$ is the Planck constant, and $c$ is the speed of light. Using $\hbar \omega = 1.3$~eV and $d_z/e = 6$~\AA~\cite{Magorrian2016}, we obtain $\tau_r \approx 3$~ns in a reasonable agreement with experimental values of $\tau_1$.

As shown in Fig.~\ref{TRPL} (b), for 11 and 16 ML, the PL is dominated by the “slow” component. However, starting from 20 ML the relative contribution of the “fast” component ($I_1/I$) to PL significantly increases, up to 20~\% in the 75-ML flake. Since the “fast” component is attributed to the band-to-band radiative recombination, such an increase indicates the transition to the direct band gap InSe, which takes place in the region of 16-20 ML. Note that, however, the defect-assisted contribution still remains large, even in the flakes with the direct band gap.

\section{Discussion}

\begin{figure}[H]
\centering
    \includegraphics[width=0.45\textwidth]{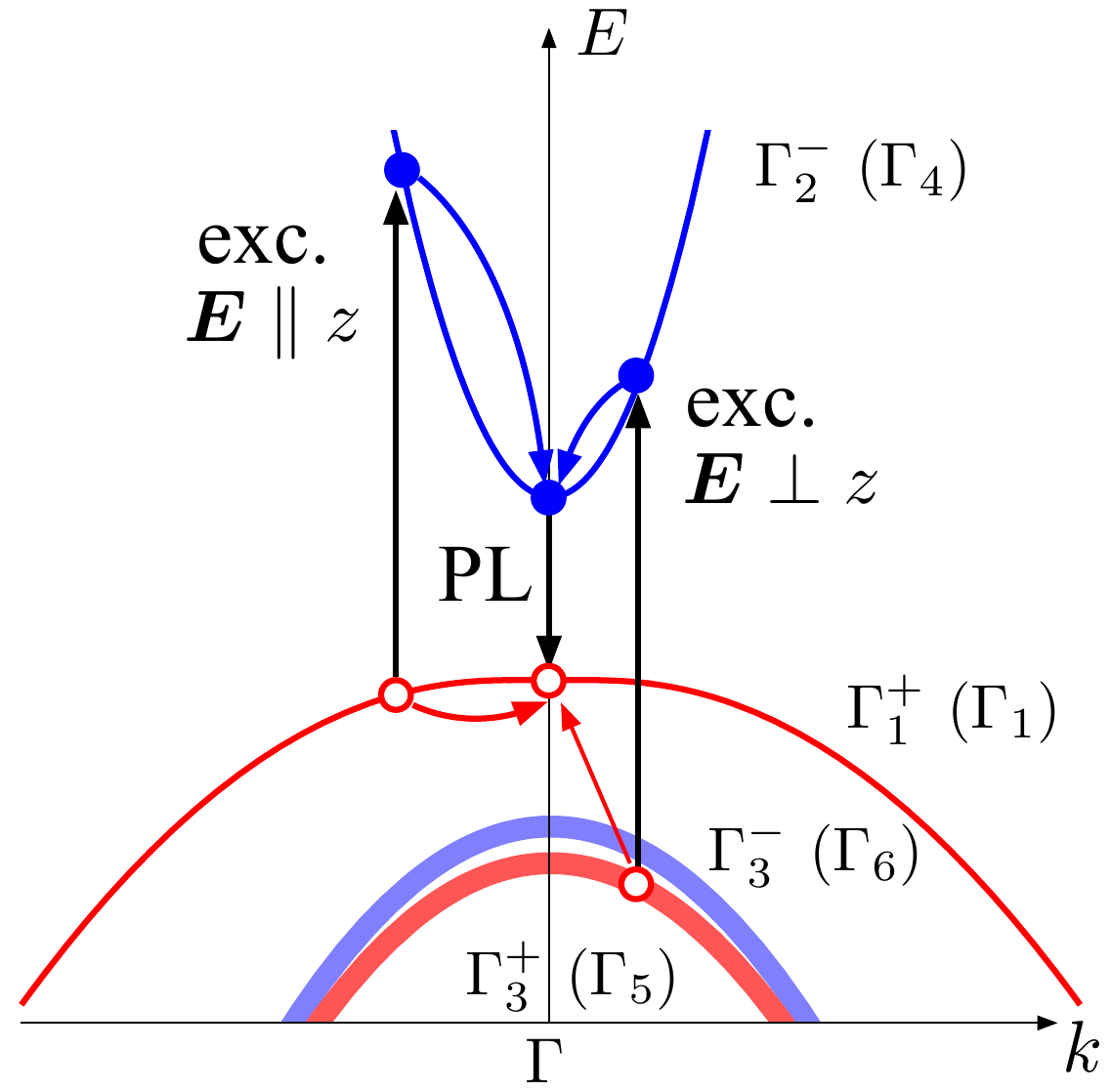}
    \caption{The scheme of excitation, relaxation and emission processes in a multilayer direct-band InSe at high excitation energy. The black arrows show optical transitions with absorption or emission of light, the blue and red arrows show the energy relaxation processes in the conduction and valence bands, respectively. $\Gamma_j^\pm$ and $\Gamma_j$ denote the irreducible representations of the $D_{3d}$ (even number of monolayers) and $D_{3h}$ (odd number of monolayers) point groups, respectively, which describe the shown bands at the $\Gamma$ point.}
    \label{model_high_energy}
\end{figure}

Let us now discuss the optical transitions and electron/hole relaxation in a few-layer InSe, which sheds light on the appearance of the  $I_2$ with a longer decay time. 
Bulk $\beta$-InSe is described by the point symmetry group $D_{6h}$, whereas the symmetry of a few-layer structure depends on the number of layers. Flakes with even layer thickness belong to the centrosymmetric $D_{3d}$ group, while flakes with odd layer thickness belong to the noncentrosymmetric $D_{3h}$ group~\cite{Li2013, rybkovskiy2014transition, Magorrian2016, Shubina2019, Sun2022}. At $\Gamma$ point of the Brillouin zone, the electronic states of the topmost valence band ($v$) transform as scalar functions corresponding to the $\Gamma_1^+$ and $\Gamma_1$ irreducible representations in the even-layer and odd-layer structures, respectively. The states of the lowest conduction band ($c$) transform as the $z$-component of vector, i.e. according to the $\Gamma_2^-$ and $\Gamma_4$ representations~\cite{Kuroda:1980,rybkovskiy2014transition, Magorrian2016, Shubina2019}, see Fig.~\ref{model_high_energy}. Therefore, the electric dipole transitions between the $v$ and $c$ bands are possible for electric field polarized in the $z$-direction, $\bf E \parallel z$. Note that spin-orbit coupling also allows for weak $c \to v$ transitions in the $x$ and $y$ polarizations.~\cite{Kuroda:1980}
The valence bands with lower energies, $v1$ and $v2$, are formed by the states that transform as $x \pm \mathrm{i} y$ and $z (x \pm \mathrm{i} y)$, respectively. The corresponding representations of the $D_{3d}$ (even $N$) and $D_{3h}$ (odd $N$) groups are $\Gamma_3^-$ and $\Gamma_6$ for the $v1$ valence band, and $\Gamma_3^+$ and $\Gamma_5$ for the $v2$ valence band, respectively~\cite{rybkovskiy2014transition, Magorrian2016}.  Electro-dipole optical transitions are allowed between $v2$ and $c$ bands for electric field lying in the plane of the layer, $\bf E \perp z$, see Fig.~\ref{model_high_energy}.

Figure~\ref{model_high_energy} schematically shows the band structure and excitation and relaxation processes in direct-band-gap InSe flakes. In our TRPL experiment the excitation photon energy is approximately 3~eV, which is high enough to excite both the $v \to c$ and $v2 \to c$ transitions (also referred to as A and B lines, respectively~\cite{Bandurin:2017,Magorrian2016}). The kinetics of relaxation is different for these two transitions. For the A transition, photoexcited electrons and holes rapidly relax to the band extrema, where they recombine radiatively with the rate $\tau_1^{-1}$. This process governs the “fast” decay component of PL shown in Fig.~\ref{TRPL}. In the case of the B transition, the photoexcited hole resides in the lower $v2$ band and, in order to recombine with an electron, should at first relax to the topmost valence band. This interband energy relaxation is slow, since it requires phonon-assisted transitions between different valence subbands, and hence, governs the corresponding PL decay time.
This hole relaxation might be responsible for the
the “medium” decay component in Fig.~\ref{TRPL} and the corresponding relaxation time $\tau_2$, although this assumption requires further investigation. 

It should be noted that any multi-exponential fitting cannot give precise boundaries between the areas of action of various mechanisms that interpenetrate, overlap, and influence each other. Therefore, our consideration is of a qualitative nature, although the main observation of the band structure transformation from indirect to direct in a certain thickness range (16–20 ML) is beyond doubt. 

\section{Conclusions}
To conclude, we have experimentally investigated the optical properties of a few-layer $\beta$-InSe at cryogenic temperature. 
Using exfoliation technique we have prepared flakes with different thicknesses varying in the 6 to 75-ML range. Raman spectra of the bulk InSe crystal used for exfoliation have shown that the studied sample has the $\beta$-type crystal structure. Kelvin probe force microscopy has revealed rapid jumps of the work function between the 16-ML and 20-ML flakes indicating a possible transformation of the electronic structure. We have observed a rapid, more than an order of magnitude, increase of the PL intensity in the same 16-20 ML interval. We attribute these rapid changes of the work function and PL signal to the transformation of the valence-band configuration from the “Mexican hat” to the parabolic one in a 20-ML flake. To uncover the PL mechanisms in our samples, we have performed the time-resolved PL spectroscopy. We have found, that 
for all the studied thicknesses, the best fit for the PL decay curve was achieved using three exponential components. 
The “fast” component is characterized by the decay time $\tau_1 \sim 0.5-1$~ns, the “medium” component decays at the time scale of $\tau_2 \sim 4-11$~ns, and the “slow” component decays in $\tau_3 \sim 40-60$~ns.
By analyzing the thickness dependence of the relative intensities of these three components and the values of the decay times, we have attributed the “fast” contribution to band-to-band recombination of electrons and holes. The “slow” component is a mixture of an indirect interband transition, when there is a band structure of the “Mexican hat” type, and defect-assisted radiation, whose contribution is significant even in bulk. Consideration of the excitation, relaxation, and recombination of carriers, characteristic of the specific band structure of InSe, based on the group theory  allowed us to establish the mechanism for the appearance of an “medium” component in the PL decay.
We believe that our results have important implications for the possible fabrication of InSe-based devices.

\section{Methods}

\subsection{Fabrication of samples}

Post-transition metal chalcogenides can be peeled off by breaking the van der Waals force between adjacent layers. The first example of mechanical exfoliation is a single-layer graphene exfoliated from graphite\cite{Geim}. 
For sample preparation, mechanical micro-cleavage and the so-called adhesive tape method were used. This method consists in sticking the tape on the surface of the material and removing the top layers. Then the tape is folded in half several times. Each time the flakes broke up into thinner parts. Finally, very thin flakes are attached to the PDMS film, which are easily transferred to a suitable substrate using the HQ Graphene 2D crystal transfer system. For simplified transfer, a slight heating of the substrate is required, about 35$^\circ$C, which this system easily provides. Flakes consisting of several monolayers were prepared by this technique using bulk $\beta$-InSe of good structural quality. Silicon wafers coated with gold 50 nm thick were used as substrates. In addition, few-layer $\beta$-InSe on Si/SiO$_2$ substrates were studied as reference samples.

\subsection{AFM characterization}
The AFM study was carried out on an Ntegra Aura scanning probe microscope (NT-MDT, Russia). To obtain the topography of the samples, high-precision probes HA\_NC (NT-MDT, Russia) with a tip curvature radius of less than 10 nm, a resonant frequency of 235 kHz, and a force constant of 12 N/m were used. The surface potential was studied using conductive probes HA\_FM/W2C+ (NT-MDT, Russia) with a tip curvature radius of 35 nm, a resonant frequency of 114 kHz, and a force constant of 6 N/m. 
The surface potential measured in the KPFM method is the difference between the work function of the probe and the sample (i.e., contact potential difference):
\[V_{CPD}^{Sample} = (\phi_{Tip} - \phi_{Sample})/e\]
To define the work function of an investigated sample, a reference sample (graphene on SiC\cite{panchal2013standardization} or HOPG\cite{hansen2001standard}) with a known work function is used. Thus, it is possible to eliminate the volatile work function of the tip and find the work function of the investigated sample using the following equation.
\[\phi_{Sample} = \phi_{Ref} + e(V_{CPD}^{Ref}-V_{CPD}^{Sample})\]
The approach is detailed in the work\cite{borodin2019kelvin}.

\subsection{Raman spectroscopy}

For structural characterization of the samples, Raman spectroscopy was used. Micro-Raman measurements were carried out using a Horiba LabRAM HREvo UV-VIS-NIR-Open spectrometer equipped with a confocal microscope. The measurements were performed at room temperature with continuous-wave (cw) excitation using the 532 nm laser line of a Nd:YAG laser (Laser Quantum Torus). An Olympus MPLN100$\times$ objective lens (NA = 0.9) was used for Raman and PL measurements, which allowed us to obtain information from an area with a diameter of $\sim1$ $\mu$m. To prevent damage to the structures, the incident laser power was limited to 0.4 mW.

\subsection{$\mu$-PL}
$\mu$-PL setup was used for optical properties investigation of InSe van der Waals structures. The sample was mounted in a He-flow cryostat ST-500-Attocube with a XYZ piezo-driver inside, which allowed us to optimize and precisely maintain the positioning of a chosen place on the sample with respect to a laser spot during a long time (few hours). Non resonant optical excitation of a cw Ti:sapphire laser (710 nm) was used for the $\mu$-PL measurements. Incident radiation was focused in $\sim$ 1 $\mu$m spot on the sample by an apochromatic objective lens with NA of 0.7. The power density was $\sim$ 6 W/cm$^{2}$. The collected emission was dispersed by a 0.5m monochromator with a 600/mm grating for detection $\mu$-PL spectra at selected wavelengths. 

\subsection{Time-resolved photoluminescence}
 To measure the TRPL spectra, we used a picosecond pulsed semiconductor laser PILAS 405 nm (Advanced Laser Systems) with repetition frequency of 4 MHz to accurately determine decay times over a large time range. Superconducting single-photon detector with a time resolution of about 40 ps and the time-correlated single-photon counting system SPC-130 (Becker Hickl) were chosen for the signal detection.

\section*{Author Contributions}
Conceptualization, B.R.B. and M.V.R.; Investigation, B.R.B, I.A.E., A.I.G., L.V.K. and M.V.R.; Formal Analysis, M.V.D. and T.V.S.; Writing – original draft, B.R.B.; Writing – review and editing, M.V.D., T.V.S. and M.V.R.; Project administration and Supervision, M.V.R. All authors have read and agreed to the published version of the manuscript.

\section*{Conflicts of interest}
There are no conflicts to declare.

\section*{Acknowledgements}
The work of B.R.B., A.I.G., I.A.E. and M.V.R. was supported by a grant from the Russian Science Foundation (no. 22-22-20049, https://rscf.ru/project/22-22-20049/) and a grant from the St. Petersburg Science Foundation in accordance with agreement no. 21/2022 dated April 14, 2022.



\balance


\bibliography{rsc} 
\bibliographystyle{rsc} 

\end{document}